\documentclass[prl,showpacs,footinbib,twocolumn,superscriptaddress]{revtex4}
\usepackage{graphics,graphicx}

\begin{document}

\title{Out of equilibrium statistical ensembles for mesoscopic rings coupled to reservoirs}

\author{S.~Peysson}
\affiliation{ITFA, Universiteit van
Amsterdam, Valckenierstraat 65, 1018XE Amsterdam, The Netherlands}
\author{P.~Degiovanni}
\address{Laboratoire de Physique de l'ENS Lyon (CNRS UMR~5672),
46 all\'ee d'Italie, 69007 Lyon, France.}
\author{B. Dou\c{c}ot}
\affiliation{LPTHE (CNRS UMR~7589), Campus Jussieu, Tour 24-14,
4 place Jussieu, 75252 Paris Cedex 05, France}
\author{R. M\'elin}
\affiliation{CRTBT (CNRS UPR 5001), CNRS, BP 166X, 38042 Grenoble Cedex, France}


\begin{abstract}
We derive non equilibrium statistical ensembles for a ballistic
Aharonov-Bohm loop connected to several electrodes connected to
reservoirs with different chemical potentials. A striking
consequence of these non trivial ensembles is the emergence
of quantum zero point fluctuations of the persistent current
around the loop. Detailed predictions for the low frequency noise
power are given.
\end{abstract}

\maketitle

Persistent currents in isolated rings have been the subject of
detailed theoretical~\cite{Imry,Landauer}
and experimental \cite{Levy,Mohanty,Chandra,Mailly}
investigations. They are indeed a probe of the many
body energy level dependance in the external magnetic flux but
they are also sensitive to the quantum state of the ring.
Persistent currents in isolated mesoscopic networks have
been discussed theoretically in the context of a semi-classical
model in the diffusive limit~\cite{Pascaud}, and in the context of
the scattering approach for ballistic
conductors~\cite{Texier}. Observation of persistent
currents in an isolated set of connected mesoscopic diffusive
rings has been reported in recent experiments~\cite{Rabaud}.

The physics associated to persistent currents in a ring connected to external reservoirs
has first been investigated in \cite{Buttiker2} where the connection between
the spectrum of the isolated ring and its transport properties in a
two terminal geometry has been established. The resulting $h/e$ periodic
oscillations of the conductance have been observed experimentally
\cite{Webb}.
Connecting a circuit (which plays the role of the {\em system})
to external leads will also affect the
persistent current because of the electron's tunnelling
into the leads. For instance,
connection to a single lead has been considered in
\cite{Buttiker-bis} to model dissipation in the ring.
Here, we investigate deeply non-equilibrium situations
such as a two
terminal experiment with a high bias voltage.
In the non-equilibrium case, the reduced density matrix
of the ring is a non trivial one, not reducible to any equilibrium
density matrix. This effect is
non perturbative since it survives even in the limit of vanishing
coupling to the reservoirs.

\medskip

In this letter, we illustrate this idea
in the simple example of a mesoscopic ballistic non interacting
ring coupled to several electrodes that
are connected to external reservoirs of different chemical
potentials. We first determine the ring's reduced
density matrix and show that it can be used to recover transport properties
in the spirit of \cite{Buttiker2}.
Then, we show that a dramatic signature of the non equilibrium
character of the ring's reduced density matrix is the presence of
quantum zero point fluctuations of the persistent current around the
ring. With progresses of low noise measurements and
nano-fabrication techniques, these
fluctuations may become observable in a near future.
This effect is basically due to the increase of electron states
that contribute to the noise. Let us mention that
increases of shot noise based on the same general idea
have also been discussed in the context of beam experiments \cite{imry:beams} and
of diffusive contacts with strong electron/electron interactions
\cite{nagaev}.
We end this letter by discussing how these
non-equilibrium issues can be traced within the framework of Keldysh's
formalism.

\medskip

\begin{figure}
\includegraphics[scale=1]{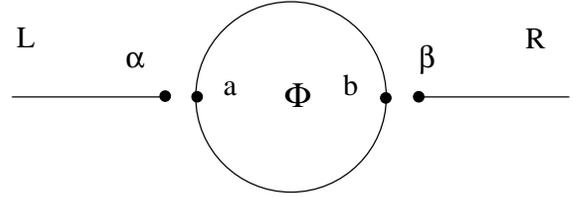}
\caption{Two terminal geometry for a mesoscopic
Aharonov-Bohm ring threaded by a magnetic flux $\Phi$.}
\label{fig:model}
\end{figure}

Similarly to Ref.~\cite{Texier}, we divide a connected system into
two types of regions: a ballistic ``intermediate'' region that
contains an Aharonov-Bohm loop and external electrodes connected
to the Aharonov-Bohm loop~(see Fig.~\ref{fig:model}). DC-transport
can be controlled through a voltage difference $V$ applied
between the left and right external electrodes and the flux
enclosed by the Aharonov-Bohm loop. Quantities of physical
interest are the transport current flowing across the mesoscopic
device and the persistent current flowing around the loop.

%%%%%%%%%%%%%%%%%%%%%%%%%%%%%%%%%%%%%%%%%%%%%%%%%%%%%%%%
%%%%%%%%%%%%%%%%%%%%%%%%%%%%%%%%%%%%%%%%%%%%%%%%%%%%%%%%

The external electrodes and the intermediate region can be
described using a tight-binding Hamiltonian for free electrons.
The coupling between the left/right electrodes and the
intermediate island is described by the tunnel Hamiltonian~:
\begin{equation}
\label{eq:tunnel} \hat{\cal W} = t_{a,\alpha} c_a^+ c_\alpha +
t_{\alpha,a} c_\alpha^+ c_a +t_{b,\beta} c_b^+ c_\beta
+t_{\beta,b} c_\beta^+ c_b.
\end{equation}
Since we are interested in out of equilibrium effects, we focus on
the $eV\gg k_BT$ regime. In most cases, we shall set $T=~0$~K.
This model is valid provided that the tunnelling amplitudes
from the electrodes to the small system are much larger than the
inelastic scattering rates inside the small system. In this case,
tunnelling to the electrodes completely determine the non
equilibrium state of the system and reservoirs. The most
transparent picture of this stationary non equilibrium state can
be obtained within the Laudauer formalism. The idea of the
scattering approach is to find a basis of stationary one-particle
states (the ``in'' states) from which the exact non equilibrium
many body state of the global system (mesoscopic device and leads)
can be constructed. The prescription is then to fill the ``in'' states
coming from each lead up to its chemical potential.

Since we are interested in the persistent current, a quantity
specific to the small system, the natural object to consider is
the one particule reduced density matrix restricted to the
mesoscopic device. To our knowledge, this
object has seldom been considered within the context of mesoscopic
physics whereas it has been discussed long ago in the context of black hole
physics. In the discussion of Unruh and Hawking effects \cite{Hawking},
the degrees of freedom behind the horizon are the counterparts of
those attached to the leads in the present case.
The ring's one particle reduced density operator can be
obtained by eliminating the one particle wave functions in the
leads from the exact stationary Schr\H{o}dinger equation. This
procedure yields an inhomogeneous Schr\H{o}dinger equation on the
small system, with an effective Hamiltonian containing
boundary terms representing hopping back and forth to the
reservoirs and a source term associated to the incident electronic
flux. Solutions to this equation provides the decomposition of
``in'' states's wave functions in the mesoscopic system's Hilbert
space.

This decomposition has a very simple form in the limit of very
small coupling to the leads. It shows Lorentzian resonances when
the energy of the ``in'' state coincides with the one particle
eigenvalue $\varepsilon_\lambda$ of the isolated mesoscopic system's
Hamiltonian. Near resonance, the decomposition of the ``in'' state
coming from the left electrode at energy $\varepsilon$ is given by:
\begin{equation}
\psi_\lambda(\varepsilon,L)\simeq\frac{i\,t_{\alpha,a}\, \langle
\lambda|a\rangle}{(\varepsilon-\varepsilon_\lambda)+\frac{i\hbar}{2}\Gamma_\lambda}.
\end{equation}
Here $\Gamma_\lambda=\frac{l}{\hbar^2v_R}(|t_{\alpha,a}\langle
a|\lambda\rangle|^2+|t_{\beta,b}\langle b|\lambda\rangle|^2)$
denotes the total escape rate given by the Fermi golden rule for
an electron in the small system stationary state
$|\lambda\rangle$ expressed in terms Fermi velocity in the
reservoirs $v_R$ and the lattice spacing $l$.
From the lead's point of view, these
resonances appear in the transmission coefficient:
\begin{equation}
t(\varepsilon)\simeq\frac{i\,t_{\alpha,a}t_{\beta,b}\,\langle b|\lambda
\rangle \langle
\lambda|a\rangle}{(\varepsilon-\varepsilon_\lambda)+\frac{i\hbar}{2}\Gamma_\lambda}.
\end{equation}
In the case where the level spacing in the mesoscopic system is
large compared to the total escape rate $\Gamma_\lambda$ of an
electron from the state $|\lambda\rangle$ to the electrodes, the
ring's reduced density matrix is diagonal within the basis of one
particle energy levels of the ring (incoherent reduced density
matrix). Populations $\overline{n}(\varepsilon_\lambda)=\langle
c^\dagger_\lambda\,c_\lambda\rangle$ are given at zero temperature
by:
\begin{equation}
\label{eq:populations}
\overline{n}(\varepsilon_\lambda)=
\frac{1}{2}-\frac{\Gamma_L}{\pi\Gamma}
\arctan{\left(\frac{\varepsilon_\lambda-\mu_L}{\hbar\Gamma/2}\right)}-\frac{\Gamma_R}{\pi\Gamma}
\arctan{\left(\frac{\varepsilon_\lambda-\mu_R}{\hbar\Gamma/2}\right)}
\end{equation}
where $\Gamma_{L,R}$ denote the escape rates in the left (resp.
right) electrode given by the Fermi golden rule. In general, these
escape rates do depend on the energy level $\lambda$. Note that
$\overline{n}(\varepsilon_\lambda)$ is nothing but the average of
contributions of all electrodes connected to the small system, the
ponderation being provided by ratios of Fermi golden rule's escape
rates. In the limit of very small tunnelling amplitudes, this
distribution can still remain non trivial, being a step function
determined by the ratios of tunnelling amplitudes. The values of
these non equilibrium populations are easily understood in a
classical way according to the sequential tunnelling picture. For
instance, when $\mu_L>\mu_R$ and for energy levels
$\mu_R<\varepsilon_\lambda <\mu_L$, the current
$(1-\overline{n}(\varepsilon_\lambda))\Gamma_L$ from the left lead
must be equal to the current
$\overline{n}(\varepsilon_\lambda)\Gamma_R$ to the right lead
leading to a non equilibrium population $\Gamma_L/\Gamma$
consistent with eq. (\ref{eq:populations}). We note that local non
equilibrium occupation numbers have been discussed and observed
experimentally in mesoscopic diffusive wires \cite{Pothier}.

For a non interacting system, the persistent current is the average
value of the derivative of the ring's Hamiltonian with respect to
the magnetic flux. Using the mesoscopic system's reduced density
matrix and under the hypothesis of negligible variation of escape
rates for all the energy levels which are partially occupied, a
simple expression can be given in terms of the persistent current
$I(\phi,\mu_\alpha)$ of a ring at fixed chemical potentiel
$\mu_\alpha$:
\begin{equation}
\label{eq:seq:perm}
I_{P}(\phi)=\sum_{\alpha}\frac{\Gamma_\alpha}{\Gamma}\,I(\phi,\mu_\alpha)
\end{equation}
The current $I(\phi,\mu_\alpha)$ can be computed directly
in term of the single electron eigenenergies $\varepsilon_n(\phi)$ for the isolated ring as
$\sum_n\frac{d\varepsilon_n(\phi)}{d\phi}\,\Theta(\mu_\alpha-\varepsilon_n(\phi))$
or equivalently using the scattering matrix of the ring
\cite{Akkermans}. Formula (\ref{eq:seq:perm}) directly shows the
influence of non equilibrium populations given by eq.
(\ref{eq:populations}). Let us also notice that, in this non
equilibrium situation, the persistent current cannot be derived
using the derivative of the average energy with respect to the
flux. As a function of the external magnetic flux, the persistent
current still roughly has a sawtooth shape but with
discontinuities when one particle energy levels cross the
reservoir's chemical potentials.

At fixed reservoir's chemical potentials, the experimental signal
is given by the Fourier transform with respect to the magnetic
flux. Let us recall that for an isolated ballistic 1D ring, the
$n$-th harmonic $I_n$ is given in terms of the Fermi velocity
$v_F$, the ring's perimeter $L$ and the number of electrons in the
ring $N$. Introducing $I_\star =ev_F/L$:
\begin{equation}
I_n=I_\star\times\frac{i}{\pi n}\,(-1)^{nN}
\end{equation}
In the non equilibrium situation described here, at zero
temperature, the harmonics are given by~:
\begin{equation}
I_n[(\mu_\alpha)]=I_\star\times\frac{i}{\pi
n}\,\times
\left(\sum_\alpha\frac{\Gamma_\alpha}{\Gamma}\,
\cos{(2\pi n\chi(\mu_\alpha))}\right)
\end{equation}
where $\chi(\mu)=\sqrt{\frac{2mL^2\mu}{h^2}}$ corresponds to an
effective flux associated with the chemical potential $\mu$.
In the particular case of a two terminal geometry, the $n$th
harmonic is modulated by the bias voltage $V$. As expected
on physical grounds, the modulation only
appears in the asymmetric case. Assuming that
$eV=\mu_L-\mu_R$ is small compared to $\sqrt{\mu} =(\sqrt{\mu_L}+\sqrt{\mu_R})/2$,
we have for the $V$-dependant part of $n$-th Fourier harmonic:
\begin{equation}
\frac{I_n[\mu_L,\mu_R]}{I_n[\mu,\mu]}-1=
\frac{\Gamma_L-\Gamma_R}{\Gamma}\, \tan{(2\pi n\chi(\mu))} \,
\sin{\left(\frac{n\,eVL}{\hbar v_F\sqrt{2}}\right)}
\end{equation}
In principle, this could be experimentally observed under
conditions similar to the ones necessary for the observation of
the persistent current in a single ballistic ring. We assume the effective
electron gas temperature to be smaller than $\hbar v_F/L$.
Tunnelling amplitudes must be within the range $k_BT/\hbar\ll\Gamma_{L/R}
\ll v_F/L$ and the voltage bias should satisfy $eV\gg\hbar
v_F/L$. For à 10~$\mu$m diameter ring, the level spacing is
typically of 500~mK. A temperature within the mK range ensures
that thermal effects do not suppress the persistent current.
Voltages above 1~mV should be sufficient to create non equilibrium
populations for many one particle energy levels. The total
dc-conductance is given by $G=\frac{e^2}{h}\frac{2L}{v_F}\,
\frac{\Gamma_L\Gamma_R}{\Gamma_L+\Gamma_R}$.

\medskip

A more striking signature of the non equilibrium reduced density
matrix can be found in the zero temperature noise of the persistent
current. Since the "in" scattering states are stationary states of
the total Hamiltonian, the two time correlation function of the
persistent current can be exactly computed using our previous
expressions for $\psi_\lambda(\varepsilon,\alpha)$. Its Fourier
transform $S_{I_P}(\omega)$ turns out to have a complicated
expression showing quantum structures above frequencies of order
$\Gamma$. For the ring considered above, at a dc-resistance of the
order of 10 to 1000 $h/e^2$, this frequency is still above 1~Mhz.
Measurements of the persistent current as carried for instance in
\cite{Rabaud} are done at a much lower frequency. Therefore,
experimentalists only access to the zero frequency limit of the
noise. $S_{I_P}(0)$ turns out to be related to the ensemble
fluctuations $C_P=\langle I_P^2\rangle -\langle I_P\rangle^2$ of
the persistent current by $S_{I_P}(0)=2C_P/\Gamma$. At low
frequency, $S_{I_P}(\omega)$ shows a lorentzian behaviour:
\begin{equation}
\label{eq:noise:lf}
S_{I_P}(\omega)\simeq \frac{2\Gamma}{\omega^2+\Gamma^2}\,C_P.
\end{equation}
The characteristic function of the probability distribution of the
persistent current is given by a superposition of partition
noises ($j_n(\phi)=d\varepsilon_n(\phi)/d\phi$):
\begin{equation}
\label{eq:perm:noise}
\widehat{P}(k)=\prod_n\left(
1+\overline{n}(\varepsilon_n(\phi))(e^{ikj_n(\phi)}-1)\right)
\end{equation}
Only the partially populated energy levels contribute to the non-zero
cumulants and therefore these non-zero cumulants constitute a
true signature of the non equilibrium state of the ring. On the
contrary, the persistent current of an isolated ring has zero noise
at zero temperature. This opens the possibility of an experimental
test although measuring the persistent current noise can be quite an experimental
challenge.

The total noise is a sum of contributions associated to the partially occupied
one particle energy levels.
For $eV\gg\hbar v_F/L$,
a continuum spectrum approximation can be used. Since all energy levels between
the smallest and the largest lead's chemical potential contribute to the
noise, the variation of the Fermi velocity $v_F(\varepsilon)$ might be taken into account for
explicit evaluations.
The variance
$C_P$ of the persistent current is finally obtained as:
\begin{equation}
C_P=I_\star^2\int \frac{v_F(\varepsilon)}{v_F}\,\overline{n}(\varepsilon)(1-\overline{n}(\varepsilon))
\,\frac{d\varepsilon}{\hbar v_F/L}
\end{equation}
In particular, for the two terminal geometry, and assuming constant escape rates and constant Fermi
velocity over the relevant energy range, one gets:
\begin{equation}
C_P=I_\star^2\,\frac{|eV|}{hv_F/L}\,\frac{\Gamma_L\Gamma_R}{\Gamma^2}
\end{equation}
which shows that the fluctuation of the persistent current becomes
larger than its average value as soon as the voltage is larger than the energy
level separation in the ring.
The three terminal geometry has a richer structure. Denoting by $\mu_{1,2,3}$ the
chemical potentials in decreasing order and $y_\alpha=\Gamma_\alpha/\Gamma$, we get:
\begin{equation}
C_P=\frac{eI_\star^2}{hv_F/L}\left(V_2y_2(y_1-y_3)+V_3y_3(1-y_3)-V_1y_1(1-y_1)\right)
\end{equation}
Note that sensitivity with respect to the intermediate voltage is enhanced with the
asymmetry in the couplings to the leads.

As expected, the noise is also a periodic function of the magnetic
flux. Its Fourier transform with respect to the magnetic flux can
easily be obtained using the free electron dispersion relation.
Using $v_F^2(\varepsilon_n(\phi))/v_F^2=
\varepsilon_n(\phi)/\varepsilon_F$, the zero temperature noise can
be related to the persistent current around a loop connected to a
single lead. Ordering the chemical potentials in increasing order,
we get:
\begin{equation}
\frac{\partial C_P}{\partial\phi}=
\frac{I_\star^2}{\varepsilon_F}\sum_\alpha
(I_P(\phi,\mu_\alpha)-I_P(\phi,\mu_{\alpha+1}))
\,
\overline{n}_{\alpha,\alpha+1}
(1-\overline{n}_{\alpha,\alpha+1})
\end{equation}
where $\overline{n}_{\alpha,\alpha+1}$ denotes the non equilibrium
population of all energy levels between chemical potentials $\mu_\alpha$
and $\mu_{\alpha+1}$.
In the end, the $n$-th harmonic of the
zero temperature variance of the persistent current is given by:
\begin{equation}
C_P(n)=\frac{I_\star^2}{2\pi^2n^2}\,
\sum_{\beta\neq\alpha}y_\alpha y_\beta\,\mathrm{sign}(\alpha-\beta)\,
\cos{(2\pi n\chi(\mu_\alpha))}
\end{equation}
Let us now connect the simple physical picture just developed to some important issues
raised within the Keldysh approach.
The Keldysh formalism is a natural framework for dealing with non
equilibrium physics \cite{Keldysh}. It provides a way to do
systematic perturbation theory in the tunnelling amplitudes. The starting
point of this perturbation theory is an ``initial'' density operator
which is partly fixed by imposing the temperatures and chemical
potentials of the external reservoirs. But this does not by itself
determine the initial density operator for the mesoscopic ring.

In order to clarify this point, let us discuss the dc-current through system. The
current flowing through the left contact can be expressed as \cite{Wingreen}:
\begin{equation}
\label{eq:ivkeldysh}
I_{\alpha,a} =
\frac{e^2|t_{\alpha,a}|^2}{\hbar}\,(G_{aa}^<g_{\alpha\alpha}^>-
G_{aa}^>g_{\alpha\alpha}^<).
\end{equation}
where the $g_{\alpha\alpha}$s denote Keldysh's Green functions for
the isolated left lead whereas the $G_{a,a}$s are for the system
with the right lead connected. Using Dyson's equation, one gets:
$G_{a,a}^<=(1+G^Rt)g^<(1+tG^A)$. Naively, equation (\ref{eq:ivkeldysh})'s
r.h.s. could
depend on the system's initial density operator through
Keldysh's Green functions for the isolated mesoscopic system. Such
terms would introduce infrared divergences in Keldysh's
perturbative expansion and of course be present in computations
involving the persistent current around the ring. But the key
point is that Dyson's equation for the retarded and advanced
Green's function provide a way to cure this problem in the non
interacting case since one can show that $1+G^Rt=G^R(g^R)^{-1}$
vanishes on the system's eigenenergies. Those terms can also be
cured order by order in Keldysh's perturbative expansion by
imposing that the system's ``initial'' Green's function $g^{>,<}$
are given by the non trivial one particle reduced density operator
computed in this letter.

Similar issues have been recently discussed in the context of quantum dots.
For example, in Ref. \cite{Coleman}, the authors ask how to represent the
effect of a large voltage bias between the two leads connected to
a dot in the Kondo regime. This particular problem raises the
important question of finding the correct starting point for
Keldysh's perturbative expansion in the coupling to the
reservoirs. In a recent work \cite{Parcollet}, O. Parcollet and C. Hooley have
pointed out the importance of finding the right reduced density
operator for the impurity spin even in the limit of a small coupling
to the leads in order to obtain a well defined and correct perturbation expansion.

\medskip
In conclusion, we have shown on a specific example that
the physics of mesoscopic devices connected to external leads
must be described using non equilibrium statistical ensembles even in the
limit of vanishing coupling to the reservoirs.
The example of a mesoscopic ring connected to two leads at different voltages
could provide a direct experimental test of these ideas. We have shown that
non zero cumulants of the persistent current are a signature of the
non-equilibrium reduced density matrix of the mesoscopic ring. These ideas can be extended to
more complicated circuits using the scattering approach on general
graphs \cite{Texier2}. Another issue is to derive the non equilibrium state
of an interacting system such as a Luttinger liquid coupled to leads.

\medskip

Fruitful discussions with S. Camalet, F. Hekking and L. Saminadayar are
acknowledged. J.-C. Cuevas is thanked for very useful comments on
Keldysh's perturbation theory.

\end{document}